\documentclass[useAMS,usenatbib]{mn2e}

\usepackage[dvips]{graphicx}
\usepackage{amsmath}
\usepackage{amssymb}

\title{Spectroscopy with the Engineering Development Array: cold H$^{+}$ at 63~MHz towards the Galactic Centre}
\author[J.~B.~R. Oonk et al.]{J.~B.~R. Oonk$^{1,2,3}$\thanks{E-mail:
oonk@astron.nl}, E. L. Alexander$^{2,4}$, J. W. Broderick$^{2}$, M. Sokolowski$^{5,6}$, R. Wayth$^{5,7}$\\ \\
$^{1}$Leiden Observatory, Leiden University, P.O. Box 9513, NL-2300 RA Leiden, The Netherlands\\
$^{2}$Netherlands Institute for Radio Astronomy (ASTRON), Oude Hoogeveensedijk 4, 7991 PD Dwingeloo, The Netherlands\\
$^{3}$SURFsara, P.O. Box 94613, 1090 GP Amsterdam, The Netherlands\\
$^{4}$Jodrell Bank Centre for Astrophysics, School of Physics and Astronomy, The University of Manchester, Manchester M13 9PL, United Kingdom\\
$^{5}$International Centre for Radio Astronomy Research, Curtin University, Bentley, WA 6102, Australia\\
$^{6}$ARC Centre of Excellence for All-sky Astrophysics (CAASTRO), Australia\\
$^{7}$ARC Centre of Excellence for All Sky Astrophysics in 3 Dimensions (ASTRO 3D), Australia\\
}
\begin{document}

\date{Accepted 0000. Received 0000; in original form 0000}

\pagerange{\pageref{firstpage}--\pageref{lastpage}} \pubyear{0000}

\maketitle

\label{firstpage}

\begin{abstract}
The Engineering Development Array (EDA) is a single test station for Square Kilometre Array (SKA) precursor technology. We have used the EDA to detect low-frequency radio recombination lines (RRLs) from the Galactic Centre region. Low-frequency RRLs are an area of interest for future low-frequency SKA work as these lines provide important information on the physical properties of the cold neutral medium. In this project we investigate the EDA, its bandpass and the radio frequency interference environment for low-frequency spectroscopy. We present line spectra from 30 to 325~MHz for the Galactic Centre region. The decrease in sensitivity for the EDA at the low end of the receiver prevents carbon and hydrogen RRLs to be detected below 40 and 60~MHz respectively. RFI strongly affects frequencies in the range 276--292, 234--270, 131--138, 95--102 and below 33~MHz. Cn$\alpha$ RRLs were detected in absorption for quantum levels \textit{n} = 378 to 550 (39 -- 121~MHz) and in emission for \textit{n} = 272 to 306 (228 -- 325~MHz). Cn$\beta$ lines were detected in absorption for \textit{n} = 387 to 696 (39 -- 225~MHz. Hn$\alpha$ RRLs were detected in emission for \textit{n} = 272 to 480 (59 -- 325~MHz). Hn$\beta$ lines were detected for \textit{n} = 387 to 453 (141 -- 225~MHz). The stacked Hn$\alpha$ detection at 63~MHz is the lowest frequency detection made for hydrogen RRLs and shows that a cold (partially) ionized medium exists along the line of sight to the Galactic Centre region. The size and velocity of this cold H$^{+}$ gas indicates that it is likely associated with the nearby Riegel-Crutcher cloud.
\end{abstract}

\begin{keywords}
ISM: clouds -- radio lines : ISM -- ISM: individual objects: Galactic centre
\end{keywords}

\section{Introduction}
The Square Kilometre Array (SKA) is a next generation telescope with unprecedented sensitivity that is currently under development. This international project consists of two telescope arrays, SKA-Mid (0.3--13.8~GHz) in South Africa and SKA-Low (50--350~MHz) in Australia. The wide variety in science objectives that will be enabled by the SKA have been detailed by the astronomical community in \citet{Br15}.

A particularly interesting science case for the SKA concerns measuring low-frequency radio recombination lines (RRL) from hydrogen and carbon and is described in \citet[][and references therein]{Oo15}. These weak lines with typical peak optical depths of 10$^{-4}$ to 10$^{-3}$ provide an important method to determine the physical conditions (density \textit{n}$_{\rm{e}}$ and temperature \textit{T}$_{\rm{e}}$) of cold, diffuse gas that are not easily obtained by other methods \citep*[e.g.][]{Oo17,Oo15,Sa18,Sa17,Sa17a,Sa17b,Ro11,Go09,Ka98a,Pa89}. Furthermore, these lines provide a largely unknown foreground signal for measurements of the HI 21~cm line from the epoch of reionization \citep[e.g.][]{Pe11,Oo15}. SKA-Low in phase 1, also known as SKA1-Low \citep{Br15}, will provide the necessary 1--2 orders of magnitude increase in surface brightness sensitivity, over existing telescopes such as the Low-Frequency Array \citep[LOFAR,][]{Ha13}, the Murchison Widefield Array \citep[MWA,][]{Ti13} and the Long Wavelength Array \citep[][]{El13}. For our Galaxy this will enable, (i) a full low-frequency Galactic plane RRL survey on arcsec to arcmin scales over a sufficiently wide frequency range to enable detailed modeling of RRL optical depths and line widths, and (ii) detailed measurements of the latitude distribution of low-frequency Galactic RRL emission \citep{Oo15}.  

The Engineering Development Array (EDA) is a single station SKA1-Low prototype system that is built on the site of the MWA in Western Australia. It is an aperture array comprising 256 dual-polarisation dipoles arranged in a pseudo-random configuration inside a diameter of 35\,m. The dipoles are MWA-style bowtie antennas that have modified front-end electronics such that they are sensitive to signals from below 50 MHz to above 300 MHz. Signals from the dipoles are combined via analogue time-delay beamforming to generate a single dual-polarisation beam from the sky. For this experiment, the signals were passed to a GPU-based spectrometer to generate high frequency resolution data, as described in Sect.~\ref{s_obs_red}. The full technical specifications are described in \citet{Wa17}. As SKA1-Low heads towards the critical design review stage in the engineering process, on-site testing and verification is essential. This includes conducting simple science experiments.

Here we report on spectroscopic observations with the EDA aimed at measuring RRLs from hydrogen and carbon. As part of this project we also characterize the spectral line sensitivity, bandpass shape and RFI environment for the EDA.

\subsection{Low-frequency RRLs \& the Galactic Centre}
RRLs are important diagnostics of the physical conditions in the interstellar medium \citep[e.g.][and references therein]{Oo17,Go09}. These lines have been observed from hydrogen (HRRL), helium (HeRRL), carbon (CRRL), sulphur (SRRL) and heavier elements (sometimes referred to as XRRL in the literature). There are two types of RRLs that can be distinguished based on their emission process: spontaneous and stimulated. RRLs arising due to spontaneous emission are often associated with discrete sources such as HII regions \citep*[e.g.][]{Pa67,Lo76,Pe78,Ro92,Ka98b,An11}. Here the gas is in local thermal equilibroum (LTE) and it predominantly emits at frequencies above 1~GHz. RRLs arising due to stimulated emission are often associated with diffuse (partially) ionized gas, such as diffuse clouds in the cold neutral medium (CNM). In this case the gas is often found not to be in LTE and it predominantly emits at frequencies below 1~GHz \citep*[e.g.][]{Sa18,Oo17,Sa17a,Sa17b,Pa89,Ko80,Sh75}.

For our test observation with the EDA we chose to observe the Galactic Centre (GC) region, as it is a bright and well known source of low-frequency RRL emission and absorption. The GC region has been observed in both HRRL and CRRL over a large range in frequency. Here we will focus on low-frequency ($\leq$1.4~GHz) RRLs from in particular hydrogen. The HRRLs measured towards the GC region are observed to be in emission at all frequencies from 141.5~MHz and upwards. Near 1.4 GHz, HRRL have been oberved and mapped with 15--30~arcmin resolution using single dish telescopes \citep*[e.g.][]{Lo73,Lo76,Ke77,Pe78,Al15}. At this high spatial resolution, the line profile shows signs of multiple components and spans a total velocity range from about -100~km~s$^{-1}$ to +80~km~s$^{-1}$ \citep[e.g.][]{Ke77}. 

At lower frequencies, HRRLs have been observed down to 141.5~MHz (\textit{n}=359) \citep*[e.g.][]{Ro00,Ro01,Ro02,Ro97,An90,An85a,An85b,Pe78}. These observations have typically been done with large, degree scale beam widths and show that on these scales there is one dominant line component at a local standard of rest (LSR) velocity \textit{v}$_{\rm{LSR}}\sim$0--10~km~s$^{-1}$, with a full width at half maximum FWHM$\sim$25~km~s$^{-1}$. This component has been referred to as the 0~km~s$^{-1}$ component by \citet[][hereafter P78]{Pe78}. P78 also show that the change in line width with spatial resolution is most likely due to beam dilution of the compact regions responsible for the broad line emission. They argue, assuming a single velocity component with a width of 25~km~s$^{-1}$, that the low-frequency HRRL component therefore most likely arises due to stimulated emision from a warm (\textit{T}$_{\rm{e}}\sim$5000--10000~K), low density (\textit{n}$_{\rm{e}}\sim$1--30~cm$^{-3}$), diffuse gas along the line of sight to the GC. Such a gas could exist as part of the warm outer envelopes of HII regions \citep[e.g.][]{An85b}. A similar conclusion was reached by \citet{An90}. \citet[][hereafter R97]{Ro97} present evidence for the possible existence of a narrower, FWHM$\sim$8~km~s$^{-1}$, extended ($\geq$1~deg) HRRL component at 328~MHz. Such a narrow line can only arise from a much colder gas with \textit{T}$_{\rm{e}}\leq$1300~K. However, R97 could not confirm this narrow feature as at this frequency it is strongly blended with the FWHM$\sim$25~km~s$^{-1}$ component and they were unable to conclusively reject radio frequency interference (RFI) as a possible origin for this feature. 

Carbon RRLs, $\alpha$ ($\Delta$\textit{n}=1) transitions, towards the GC were detected by \citet{An88} near 74~MHz in absorption at a velocity (LSR) of -1~km~s$^{-1}$. This was confirmed by the CRRL Galactic plane survey by \citet{Er95} at 76~MHz. They find extended CRRL absorption and towards the GC these have a peak line-to-continuum ratio of about 10$^{-3}$, a \textit{v}$_{\rm{LSR}}\sim$0--5~km~s$^{-1}$ and a FWHM$\sim$24~km~s$^{-1}$. At even lower frequencies, \citet*{Ka01} detect CRRL absorption at 34.5~MHz. At higher frequencies, R97 measure extended CRRLs in emission towards the GC at 328~MHz with \textit{v}$_{\rm{LSR}}\sim$0--7~km~s$^{-1}$ and a FWHM$\sim$20~km~s$^{-1}$. P78 measure CRRL emission at 408~MHz with a line width and velocity similar to that of R97. 

\section[]{Observations \& Data Reduction}\label{s_obs_red}
We used the EDA with a drift-scan method to conduct these observations. Scans were taken in 2 minute intervals, with the array being directly centered on the source (GC) in the middle of each scan. During each 2 minute scan no tracking was performed. The telescope beam was re-pointed between scans. The sampling time was 2 seconds (i.e. we obtain 60 individual spectra per scan) and the channel width was 1.25~kHz. In total 262144 channels were observed, providing a bandwidth of 327.68~MHz. To achieve this high spectral resolution over this large bandwidth it was only possible to record data for 1 polarization (the X-polarization) with the EDA GPU-based spectrometer.
The bandwidth and frequency range recorded here with the EDA is similar to SKA1-Low, which will have a 300~MHz bandwidth, from 50 to 350~MHz. Originally SKA1-Low planned to have 262144 channels\footnote{The original SKA1-Low design is described at https://astronomers.skatelescope.org/wp-content/uploads/2016/12/SKA-TEL-SKO-DD-001-1\_BaselineDesign1.pdf} with 1.14~kHz width each, but this has recently been reduced to 65536 channels\footnote{The current SKA1-Low design is described at https://astronomers.skatelescope.org/wp-content/uploads/2015/11/SKA-TEL-SKO-0000007\_SKA1\_Level\_0\_Science\_RequirementsRev02-part-1-signed.pdf} with 4.58~kHz width each\footnote{For SKA1-Low up to four zoom windows are foreseen that will allow for better frequency resolution at the expense of covering less bandwidth.}. 

\begin{figure*}
    \includegraphics[width=0.49\textwidth, angle=0]{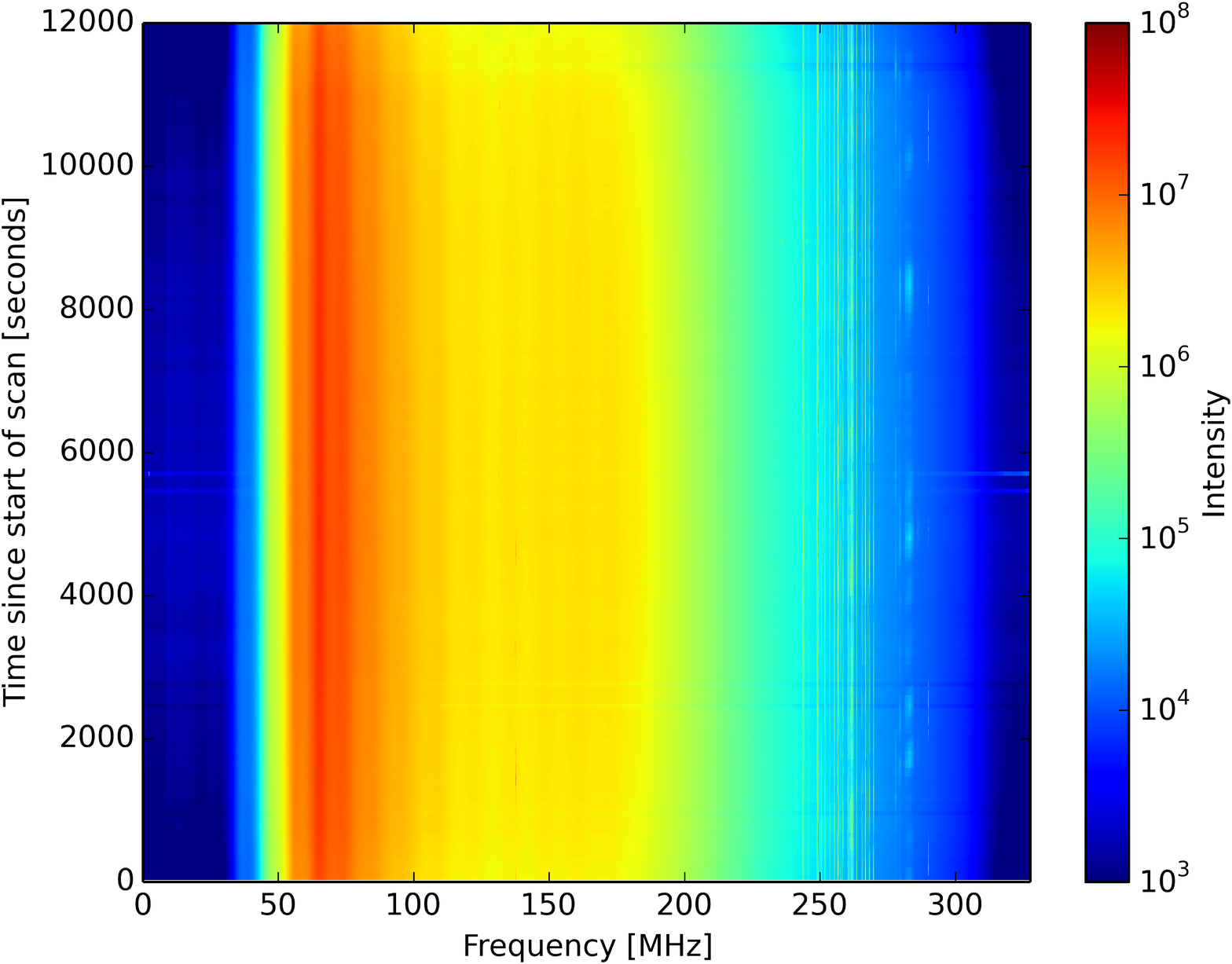}
    \includegraphics[width=0.49\textwidth, angle=0]{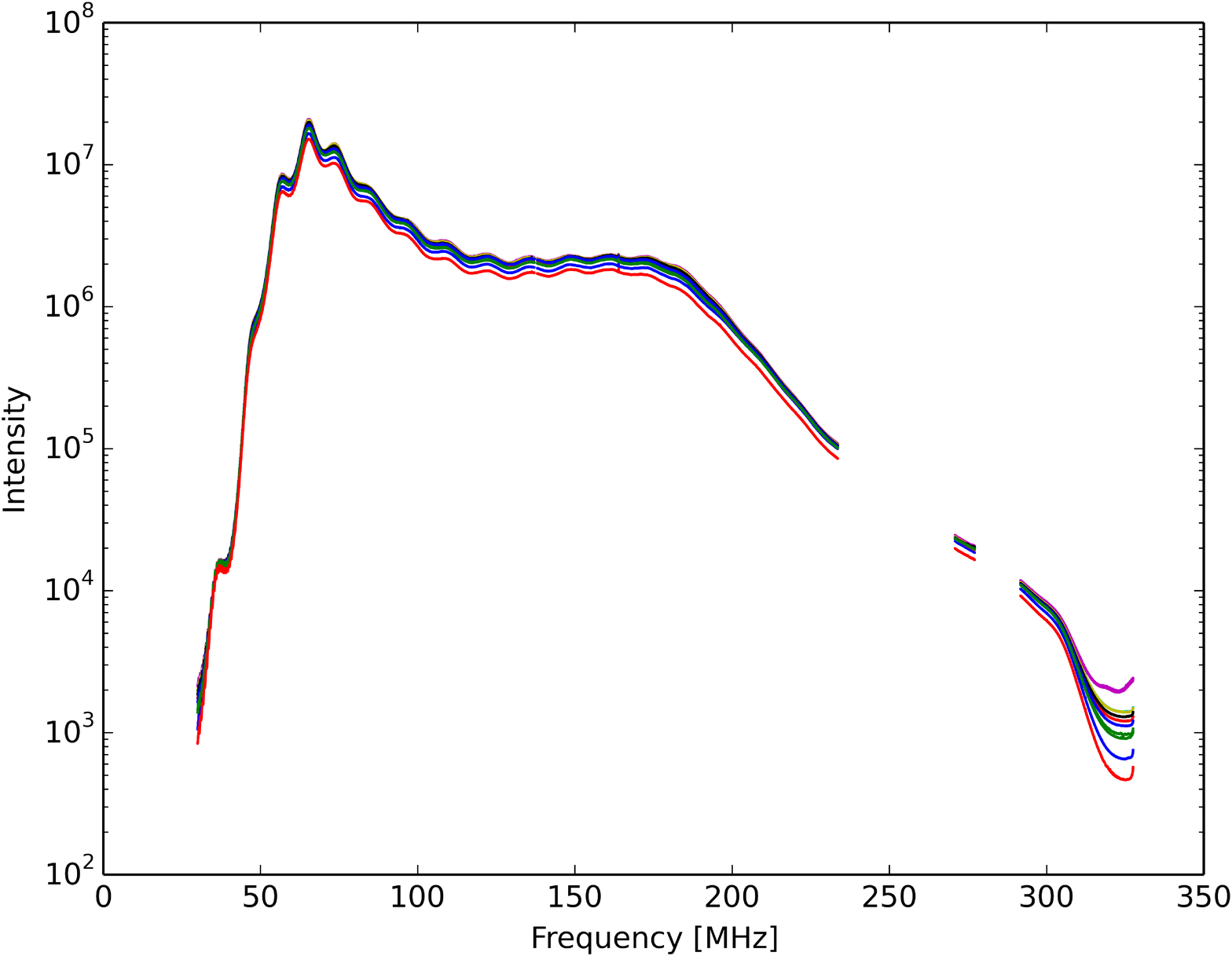}
  \vspace{0.5cm}
  \caption{EDA observation of the GC region. (Left) Observed data shown as frequency (horizontal axis) vs. time (vertical axis). (Right) Bandpass shown as the average power in arbitrary units (A.U.) vs. frequency for each 20~min scan. Different colors represent different scans and show the stability of the bandpass profile over time.}\label{f_data_bpass}
\end{figure*}

The details of the EDA observation are summarized in Table~\ref{t_obs}. The data were obtained in the form of a dynamic spectrum (i.e. time vs. frequency) and are shown in Fig.~\ref{f_data_bpass} (in arbitrary intensity units). Useful data are obtained between 30 and 325~MHz, but RFI is observed to strongly affect frequencies in the ranges 276--292, 234--270, 131--138, 95--102 and below 33 MHz. Over this frequency range the FWHM beamsize for the EDA varies from 1.5~degrees at 325~MHz to 16.4~degrees at 30~MHz. Absolute flux and bandpass calibration observations were not carried out at the time of the observation and follow-up observations could not be carried out because of activities related to the deployment of the Aperture Array Verification System \citep{Be17}. The absolute flux scale is not important as for RRLs the quantity of interest is optical depth. However, contamination from sidelobe emission may affect our optical depth measurements. At low frequencies the sky temperature dominates the system temperature. On degree scales, the GC region is the brightest source on the sky and using low-frequency continuum maps \citep*{Ha82,Do17} we estimate that any sidelobe contamination is less than 10~percent at any frequency for this EDA observation. Directly comparing our measurments near 145~MHz with \citet[][hereafter APB90]{An90}, that have the same beamsize, shows consistency between the measurements and hence that sidelobe contamination is not significant.

\begin{figure*}
    \includegraphics[width=0.8\textwidth, angle=0]{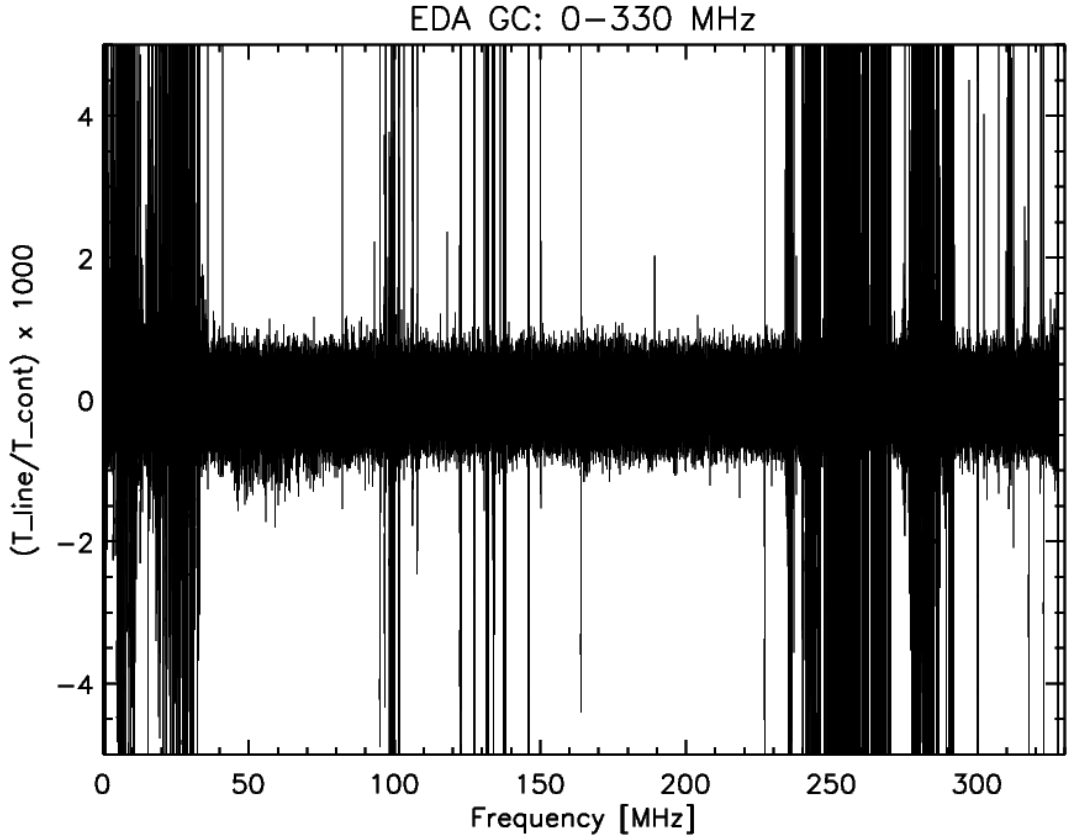}
  \vspace{0.5cm}
  \caption{EDA observed (time averaged) spectrum towards the GC region in units of $(T_{\rm{line}}/T_{\rm{cont}})\times$1000, with $T_{\rm{line}}$ the line temperature and $T_{\rm{cont}}$ the continuum temperature.. Note that in the range from 40--90~MHz the carbon RRLs are directly visible in absorption at $(T_{\rm{line}}/T_{\rm{cont}})\times$1000 around -1. The signal spikes with an absolute value of $(T_{\rm{line}}/T_{\rm{cont}})\times$1000 larger than about 2 are all associated with strong RFI. In the 30--325~MHz range and away from regions with strong RFI the 1$\sigma$ RMS noise level is reasonably constant around 3$\times$10$^{-4}$, per 1.25~kHz channel, in units of $T_{\rm{line}}/T_{\rm{cont}}$.}\label{f_fullspec_tau}
\end{figure*}

Bandpass calibration is important. The frequency response of the telescope was found to be smooth over $\sim$0.16~MHz scales. We therefore performed bandpass corrections using a 3rd order polynomial over 0.16~MHz ranges. This is sufficient for the narrow (FWHM~$\sim$~1--30~kHz) RRLs observed here. The observed dynamic spectrum shows small jumps in the signal level from scan to scan. This corresponds to the re-pointing of the array. To create a normalized spectrum for each scan, we first average the 60 data points per channel. Masking the expected RRL frequencies, we fit the bandpass across the averaged spectrum of each scan over blocks of 128 channels (0.16 MHz) in width. Following \citet{Oo14} we divide the observed scan spectrum by this bandpass and subtract one to create a bandpass corrected spectrum in dimensionless units of \textit{T}$_{\rm{line}}$/\textit{T}$_{\rm{cont}}$.\footnote{This dimensionless quantity equals negative optical depth, i.e. \textit{T}$_{\rm{line}}$/\textit{T}$_{\rm{cont}}$=-$\tau_{\rm{line}}$.} Here \textit{T}$_{\rm{line}}$ and \textit{T}$_{\rm{cont}}$ are the measured line and continuum temperatures respectively. This is repeated for each scan. The final spectrum is then created by averaging the corrected spectra for all scans, see Fig.~\ref{f_fullspec_tau}. The EDA does not provide Doppler tracking and hence we apply this correction in our offline data reduction procedures.

\section[]{Results}\label{s_results}
\begin{figure*}
\mbox{
    \includegraphics[width=0.74\textwidth, angle=90]{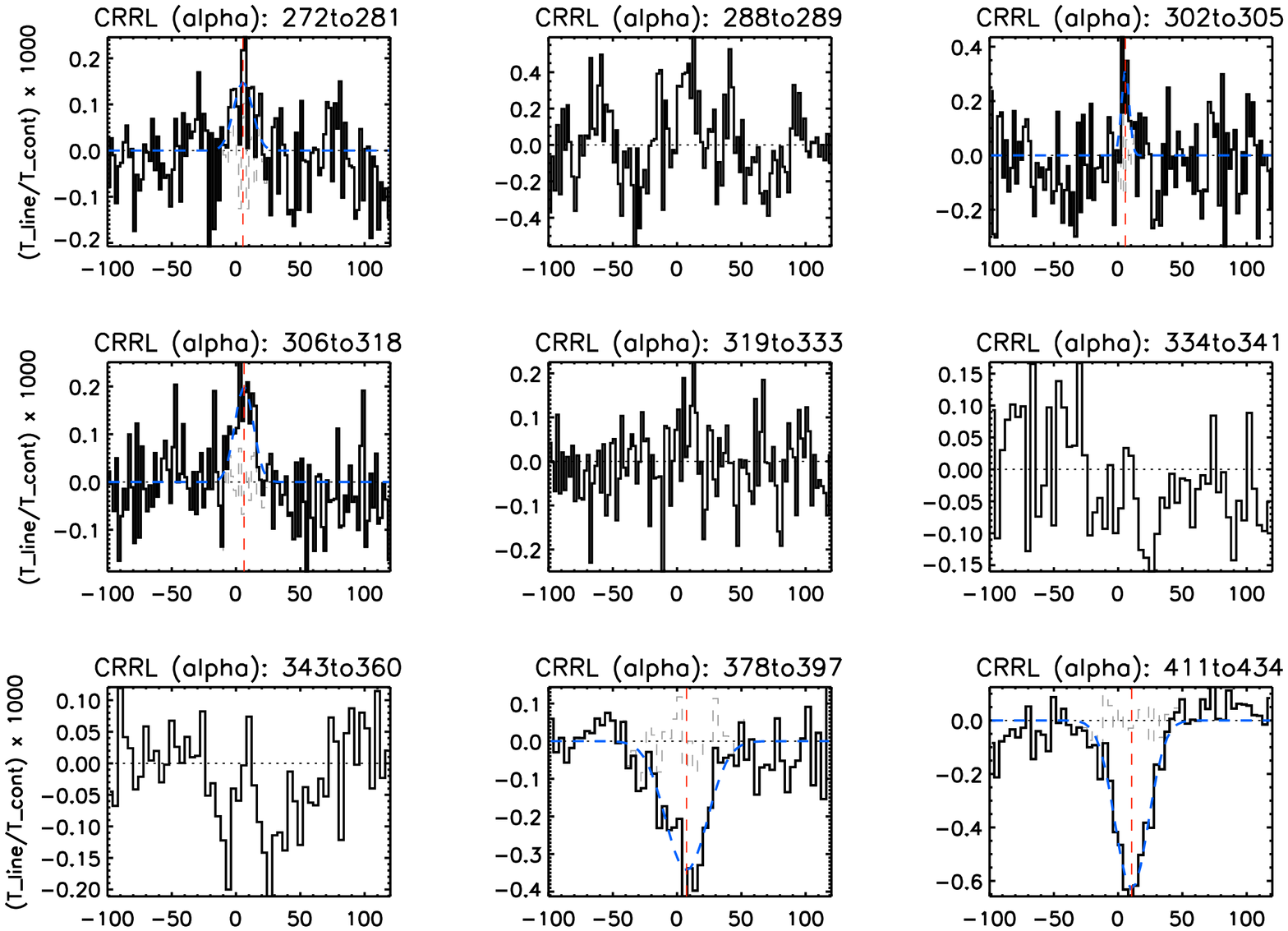}
}
\mbox{
    \includegraphics[width=0.74\textwidth, angle=90]{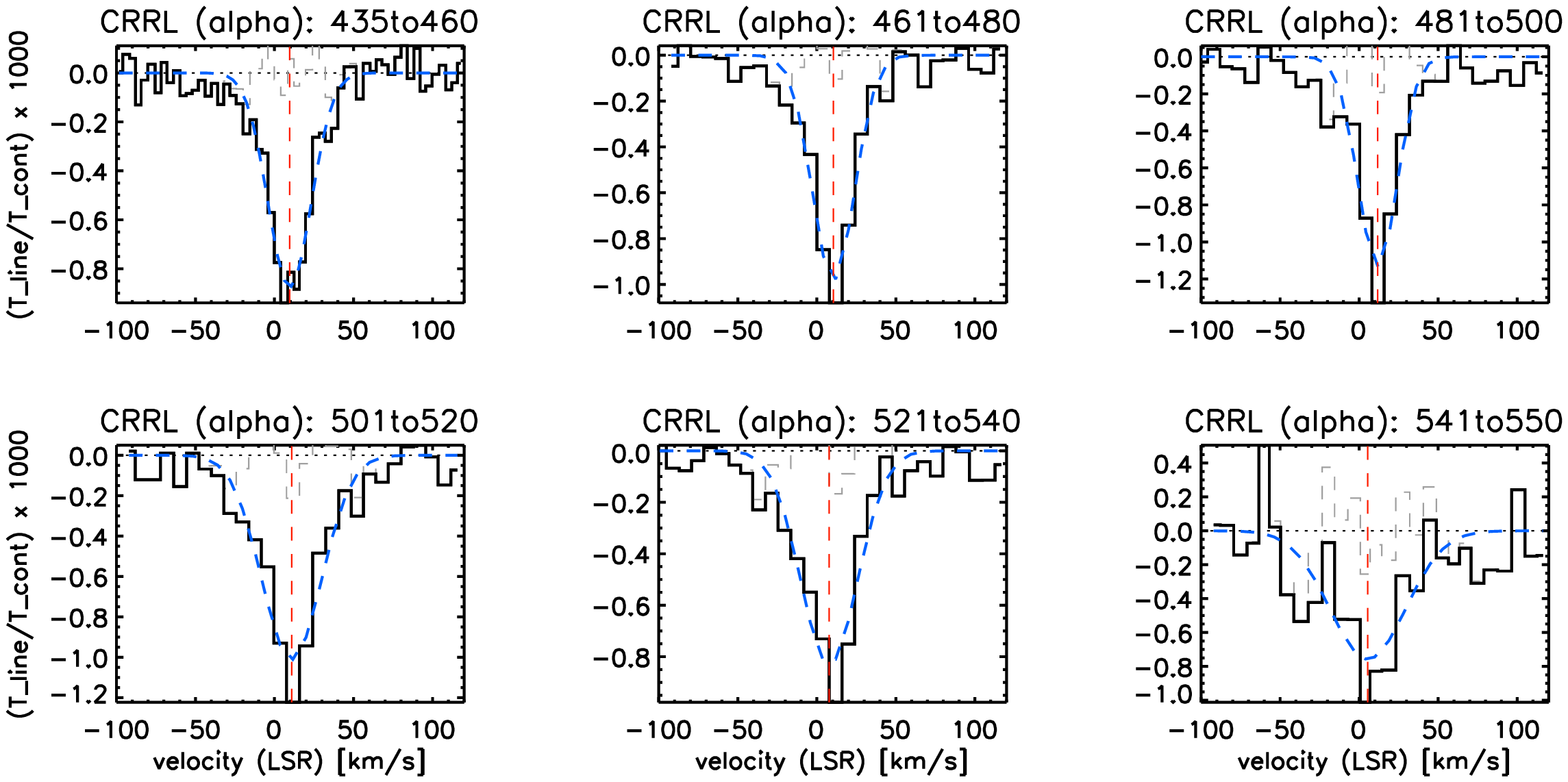}
}
  \vspace{-4.5cm}
  \caption{Cn$\alpha$ stacked RRL spectra for different n-ranges. The thick dashed (blue) line shows the Gaussian fit. The dark, thin dashed (red) line shows the fitted line centroid and the light, thin dashed (grey) line shows the fit residuals. The x-axis and y-axis labels are same for all panels.}\label{f_crrl_mplt}
\end{figure*}

C$\alpha$ RRLs were detected in absorption for \textit{n} = 378 to 550 (39--121~MHz) and in emission for \textit{n} = 272 to 318 (204--325~MHz); see Fig.~\ref{f_crrl_mplt} and Table~\ref{t_cna_rrl_stacked}. Here \textit{n} is the quantum number of the line transition. The \textit{n} value for this emission to absorption transition directly indicates that we are observing cool, diffuse gas associated with the CNM \citep[e.g.][]{Sa17b,Oo17,Pa89} and we will discuss this in more detail in Sect.~\ref{s_discuss}. In emission these lines have a narrow line profile with a FWHM of about 17~km~s$^{-1}$, whereas in absorption they have a nearly constant FWHM of about 32~km~s$^{-1}$. This change in FWHM cannot be due to pressure or radiation broadening as these scale with n$^{5.2}$ and n$^{5.8}$ respectively. It is more likely that the change in beam size is responsible for the observed change in line width. For n$\ge$500 ($<$52~MHz) an increase in the FWHM to about 56~km~s$^{-1}$ is observed. It is likely that also here the change in beam size is responsible. However, pressure or radiation broadening cannot be ruled out for n$\ge$500. The line profile shapes themselves will be discussed in more detail in Sect.~\ref{s_line_profile}. The central velocity of the C$\alpha$ lines is found to shift slightly from around 6~km~s$^{-1}$ to 10~km~s$^{-1}$ with increasing n. The integrated optical depth of the lines is rather constant for n$<$318. For n$>$378 the optical depth first increases up to n$\sim$510 before it decreases again.

\begin{figure*}
    \includegraphics[width=0.63\textwidth, angle=90]{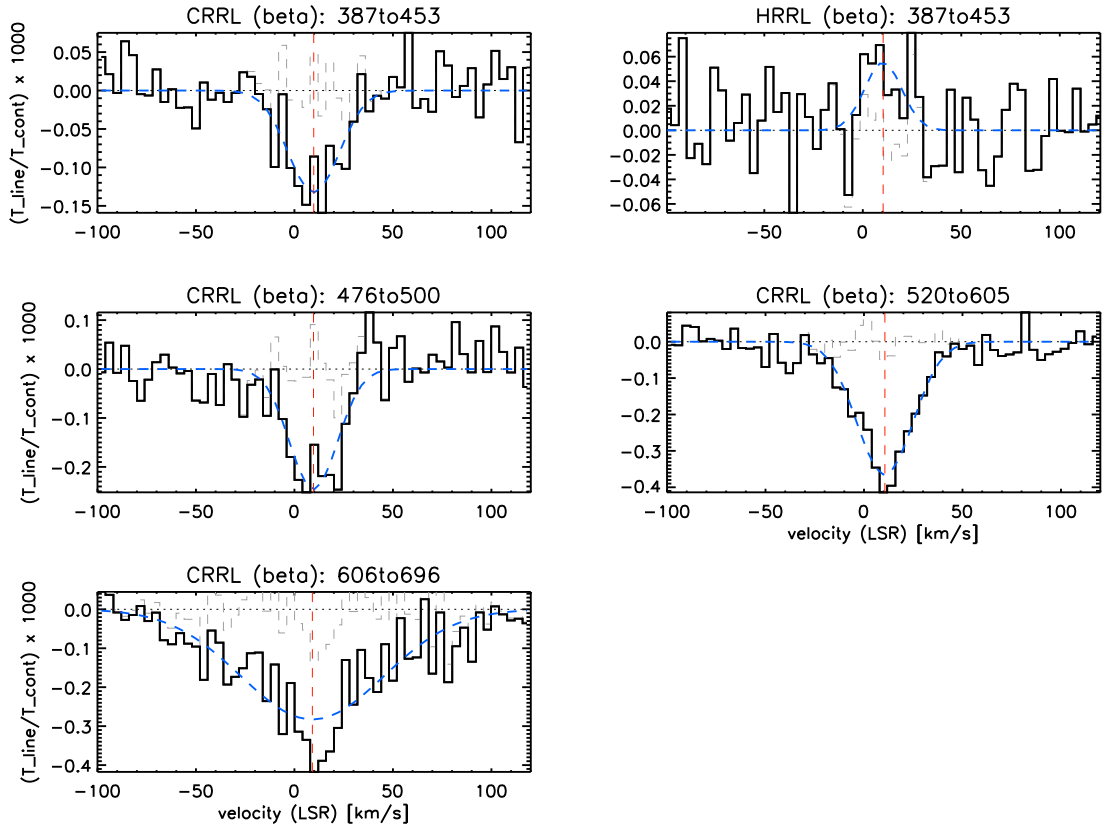}
  \vspace{0.5cm}
  \caption{Cn$\beta$ and Hn$\beta$ stacked RRL spectra. For Hn$\beta$ we have only one detection and show it in the upper right corner for direct comparison with the corresponding Cn$\beta$ stack in the upper left corner. The middle and lower panels show Cn$\beta$ detection at lower frequencies. The thick dashed (blue) line shows the Gaussian fit. The dark, thin dashed (red) line shows the fitted line centroid and the light, thin dashed (grey) line shows the fit residuals. The x-axis and y-axis labels are same for all panels.}\label{f_chbeta}
\end{figure*}

C$\beta$ ($\Delta$\textit{n}=2) RRLs were detected only in absorption for \textit{n}=387 to 696 (39--225~MHz); see Fig.~\ref{f_chbeta} and Table~\ref{t_cnb_rrl_stacked}. For \textit{n}=387--500 the C$\beta$ line widths have a FWHM of about 32~km~s$^{-1}$, in agreement with the corresponding $\alpha$ lines. For \textit{n}=520--605 the FWHM of the C$\beta$ lines is only 33~km~s$^{-1}$, almost 10~km~s$^{-1}$ less than the corresponding C$\alpha$ lines at the same n. This difference is likely due to beam dilution as at the same frequency (76~MHz) the C$\alpha$ lines do have the same width as the C$\beta$ lines. This would imply that at this frequency at least part, if not most, of the broadening is due to the increased beam sizes. At even higher \textit{n}=606--696 a strong increase in line width to a FWHM of 86~km~s$^{-1}$ is observed. This increase can be attributed to either pressure or radiation broadening, or due to the change in beam size. The central velocities of the C$\beta$ lines are around 10~km~s$^{-1}$ which agrees with the C$\alpha$ lines that are observed in absorption. The integrated optical depth of the C$\beta$ lines is found to steadily increase with \textit{n} from \textit{n}=387 to 696.

\begin{figure*}
\mbox{
    \includegraphics[width=0.63\textwidth, angle=90]{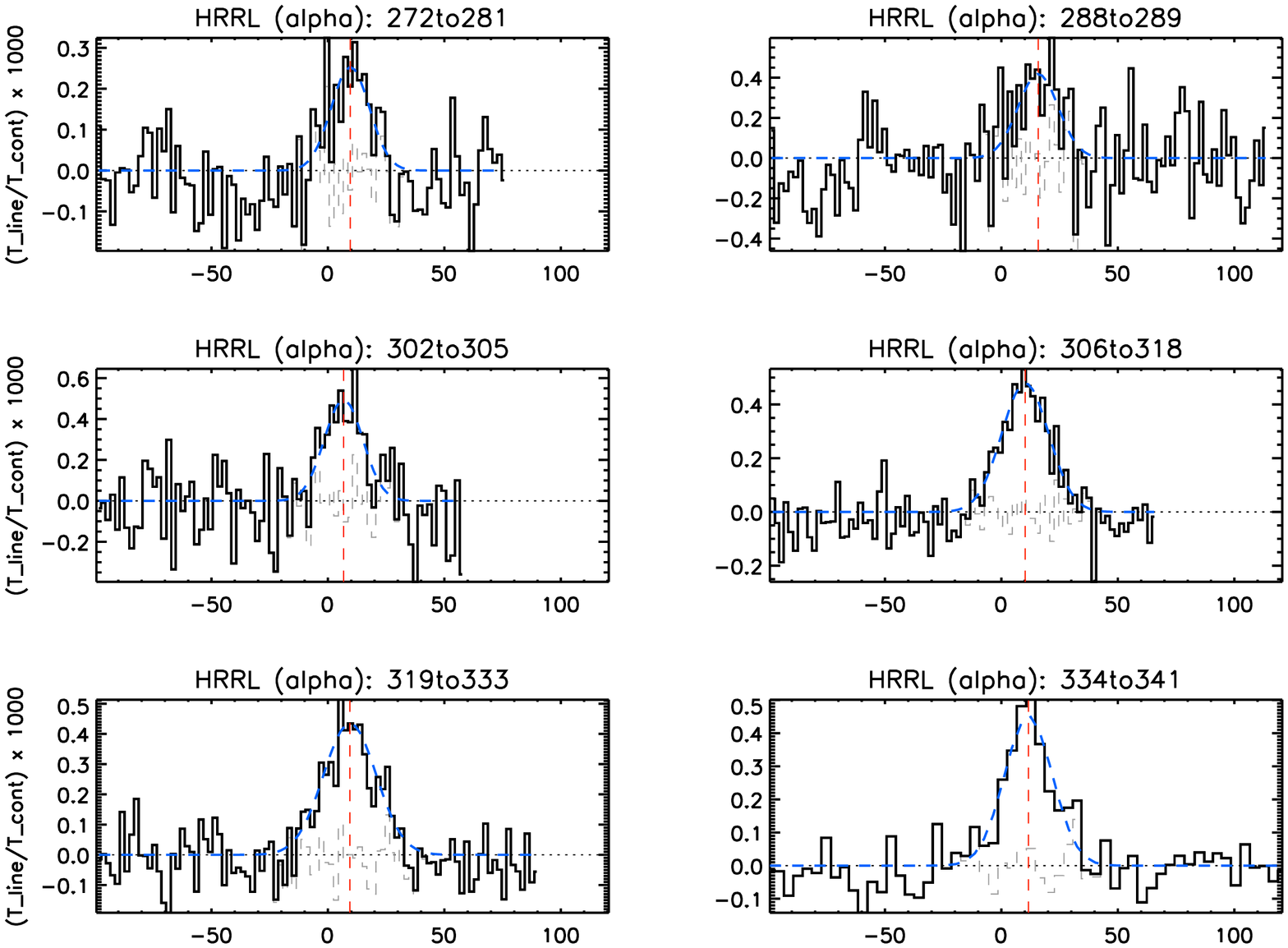}
}
\mbox{
    \includegraphics[width=0.63\textwidth, angle=90]{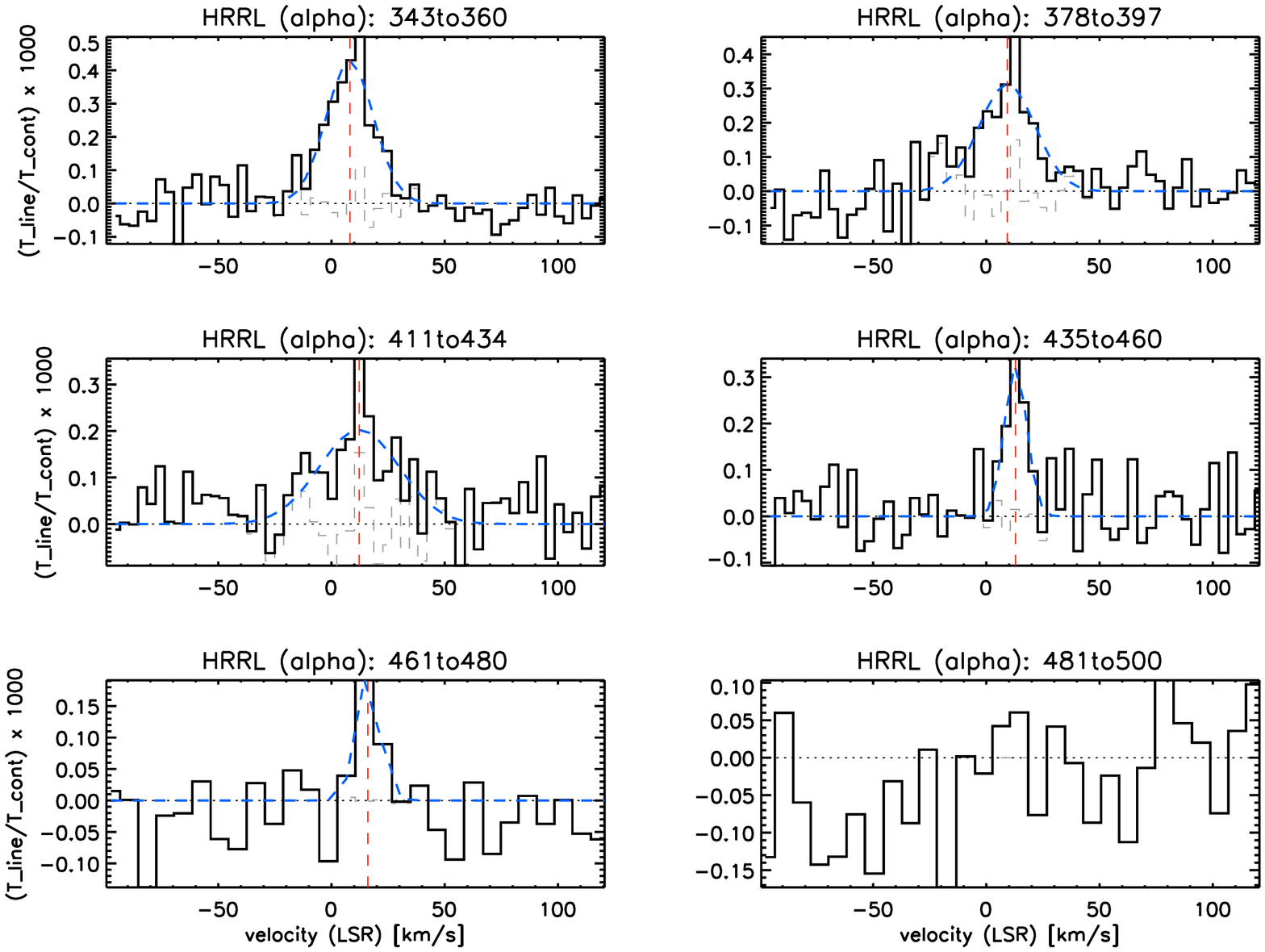}
}
  \vspace{0.5cm}
  \caption{Hn$\alpha$ stacked RRL spectra for different n-ranges. The thick dashed (blue) line shows the Gaussian fit. The dark, thin dashed (red) line shows the fitted line centroid and the light, thin dashed (grey) line shows the fit residuals. The x-axis and y-axis labels are same for all panels. For n$\ge$481 ($\leq$59 MHz) we have tried stacking Hn$\alpha$ over a variety of n-ranges, but do not detect HRRL's. As an example of these non-detections we have shown here the \textit{n}=481--500 line stack.}\label{f_hrrl_mplt}
\end{figure*}

H$\alpha$ RRLs were detected in emission for \textit{n} = 272 to 480 (59--325~MHz); see Fig.~\ref{f_hrrl_mplt} and Table~\ref{t_hna_rrl_stacked}. The line width shows a modest increase from about 20 to 30~km~s$^{-1}$ over the range \textit{n}=272--397. For n$>$435 the line width decreases sharply to around 12~km~s$^{-1}$ and is only marginally resolved with the EDA. This decrease in line width could be caused by beam dilution and we will discuss this in more detail in Sect.~\ref{s_discuss}. The central velocity of the H$\alpha$ lines is found to shift slightly from around 10~km~s$^{-1}$ to 14~km~s$^{-1}$ with increasing n. The magnitude of this shift is similar to what is found for the C$\alpha$ lines. However, the central velocity of the H$\alpha$ lines is on average 3--4~km~s$^{-1}$ higher as compared to C$\alpha$ at the same frequencies. The absolute optical depth of the line increases from \textit{n}=272 to 318, after which it steadily decreases again.

An H$\beta$ RRL was marginally detected at n$\sim$420; see Fig.~\ref{f_chbeta} and Table~\ref{t_hnb_rrl_stacked}. To verify this detection we varied the range (within \textit{n}=342 to 453) and the number of lines included in our stacking procedure. For all of our stacks, containing 39--85 lines, a 3--4$\sigma$ feature was found with an integrated optical depth varying between -0.6 and -1.0~Hz. The feature is always observed at the same central velocity of $\sim$10~km~s$^{-1}$ and has a FWHM of about 22~km~s$^{-1}$. This is in good agreement with the H$\alpha$ detections and provides additional evidence that this feature is a real H$\beta$ line.

\subsection[]{RRL $\alpha$ to $\beta$ ratio}\label{s_ab_ratio}
For gas in local thermal equilibrium (LTE) we expect the $\alpha$ to $\beta$ line optical depth ratio to be 3.6, when measured at the same frequency \citep*[e.g.][]{Wa74,Sh75,Sa17b}. Differences from this value would therefore indicate that the gas is not in LTE.  
 
Our highest signal to noise C$\beta$ \textit{n}=520--605 detection covers the frequency range 59.1--93.0~MHz. For this frequency range we have stacked the corresponding C$\alpha$ \textit{n}=411--480 lines and for it we find an integrated optical depth $\int\tau$=6.87$\pm$0.16, \textit{v}$_{\rm{LSR}}$=10.1$\pm$0.36 and FWHM=31.3$\pm$0.8~km~s$^{-1}$. The velocity and FWHM for the $\alpha$ profile match well with those of the $\beta$ profile. The integrated optical depth ratio C(411--480)$\alpha$/C(520--605)$\beta$=2.1$\pm$0.1. This is in good agreement with \citet{Er95} who measure the C(441)$\alpha$/C(555)$\beta$=2.1$\pm$0.2.This low $\alpha$ to $\beta$ ratio shows that the CRRL emitting gas is not in LTE.

Our H$\beta$ \textit{n}=387--453 detection covers the frequency range 140.6--225.3~MHz. For this frequency range we have stacked the corresponding H$\alpha$ \textit{n}=307--360 lines and for it we find $\int\tau$=7.11$\pm$0.24, \textit{v}$_{\rm{LSR}}$=9.38$\pm$0.41~km~s$^{-1}$ and FWHM=25.1$\pm$1.0~km~s$^{-1}$. The velocity and FWHM for the $\alpha$ profile match well with those of the $\beta$ profile. The integrated optical depth ratio H(307--360)$\alpha$/H(387--453)$\beta$ is 8.9$\pm$2.7. This high $\alpha$ to $\beta$ ratio shows that the HRRL emitting gas is not in LTE. This is consistent with P78 who argue that the low-frequency HRRL are out of LTE when compared with HRRL models for an optically thin cloud. At higher frequencies, 1.6~GHz, \citet*{Sh77} measure H(159)$\alpha$/H(200)$\beta\sim$4 which is consistent with LTE. This indicates that the higher frequency ($\geq$1.6~GHz) HRRL emission differs from that at lower frequencies (e.g. P78).

\citet{Sa17b} show that the $\alpha$ to $\beta$ line width ratio, when measured at the same frequency, is a useful quantity to investigate line broadening. For carbon we find that down to about 76~MHz the C$\alpha$ and C$\beta$ line widths agree. This implies that here Doppler broadening dominates. Below this frequency the C$\beta$ lines have an increased line width as compared to the $\alpha$ lines, indicating that at these frequencies pressure and radiation broadening become inportant. Similar behavior for the CRRL line widths was also found for the cool clouds along the line of sight towards Cassiopeia~A \citep[e.g.][]{Oo17,Sa17,Ka98a,Pa89}. The hydrogen line width will be discussed in more detail in Sect.~\ref{s_discuss} but our measurements at 183~MHz imply that at this frequency Doppler broadening still dominates.

\subsection[]{RRL line profiles}\label{s_line_profile}
The HRRL line profiles can be described by a single Gaussian at all frequencies. This is not true for the CRRL line profiles, which are only fitted well by a single Gaussian above 180~MHz (emission). In the 113--170~MHz range, where C$\alpha$ transitions from emission to absorption, we find dual peaked line profiles for both C$\alpha$ and C$\beta$. These dual peaks are offset from the main peak at other frequencies. This can be interpreted as evidence that the observed line profile consists of several distinct components with different physical conditions. Studies by \citet{An89} and \citet{Ro02} also show evidence for multiple velocity components. 

At lower frequencies, where the CRRLs are in absorption, the C$\alpha$ line profile starts to show signs of assymetries and broad, possibly Lorentzian, wings that may be associated with radiation and/or pressure broadening. Given that the CRRL line profile at the spatial resolution of the EDA very likely consists of several distinct physical components we do not attempt to disentangle the observed line profiles here. The H$\alpha$ lines do not show strong evidence for deviations from a Gaussian profile. The largest change observed is the decrease in FWHM from about 25~km~s$^{-1}$ to 12~km~s$^{-1}$ for n$>$435. This transition happens gradually and is most clearly observed in the \textit{n}=411--434 stack. Here a narrow line component at \textit{v}$_{\rm{LSR}}\sim$10~km~s$^{-1}$ starts to dominate and a single Gaussian fit does not fit the peak very well and thus overestimates the FWHM for this stack. Below we will focus on this narrow $\sim$12~km~s$^{-1}$ line profile, as it has important implications for the physical conditions of the associated gas.  

\subsection[]{Molecular lines}\label{s_mol_lines}
Tentative molecular line features at frequencies near 115~MHz have been reported by \citet{Tr17}. We searched our EDA data for molecular lines, using line lists from the Leiden Lambda \citep{Sc05} and the Splatalogue\footnote{http://www.cv.nrao.edu/php/splat/} catalogs. In the 30--325~MHz range and away from regions of strong RFI we find that the 1$\sigma$ RMS spectral noise in our data is nearly constant at a level of 3$\times$10$^{-4}$ in units of optical depth. In this frequency range we find no evidence for molecular lines. 

In particular we searched for individual lines from NO, OH, C$_{3}$H, H$_{2}$CO and t-HCOOH, and exclude their presence at a 3$\sigma$ peak optical depth level of 1$\times$10$^{-3}$ per 1.25~kHz channel. These lines have a multitude of transitions in the frequency range probed by our observations and we can therefore use line stacking to probe deeper. Stacking NO in the range 180--210~MHz we derive a 3$\sigma$ peak optical depth upperlimit of 2$\times$10$^{-4}$ per 1.25~kHz channel. Similarly, we exclude NO, OH, C$_{3}$H, H$_{2}$CO and t-HCOOH at a 3$\sigma$ peak optical depth upperlimit per 1.25~kHz channel of 2$\times$10$^{-4}$, 4$\times$10$^{-4}$, 3$\times$10$^{-4}$, 3$\times$10$^{-4}$ and 5$\times$10$^{-4}$, in the ranges 45--52~MHz, 87--194~MHz, 50--91~MHz, 45--75~MHz and 40--90~MHz, respectively.

If molecular line emission mainly arises due to spontaneous emission then our non-detections are not surprising as their associated Einstein A coefficients are of the order 10$^{-21}$ to 10$^{-15}$~s$^{-1}$. Alternatively, if these lines are associated with low filling factor gas, relative to the EDA beam, then such line emission would be strongly diluted and difficult to detect with the EDA.

\section[]{Discussion}\label{s_discuss}
With the EDA we have observed C$\alpha$ transitions in the range 39--325~MHz, C$\beta$ in the range 39--225~MHz, H$\alpha$ in the range 59--325~MHz and H$\beta$ near 183~MHz. The wide range over which we detect these RRLs would allow us to fit the detailed RRL models by \citet{Sa17a} to obtain the physical conditions of the emitting gas. However, we do not attempt this here given the large variation in spatial resolution with frequency for the EDA. The GC region, being an extended and complicated source, would require detailed aperture corrections to align the measurements at different frequencies. Such aperture corrections cannot be done very accurately with our current data set.

\subsection{Carbon RRLs}\label{s_discuss_crrl}
Comparing our CRRL measurements with previous studies \citep[e.g. P78; R97;][]{An88,Er95,Ka01,Sa17b} we find reasonable agreement. Our measurements indicate a typical LSR velocity between 5 and 12~km~s$^{-1}$. This is consistent with R97 and slighty higher than P78. The latter find LSR CRRL velocities in the range -2 to 7 km~s$^{-1}$. The differences in the integrated optical depths, line widths and velocity between the various CRRL investigations can likely be accounted for by the different beam sizes used. The offset in velocity, by about 3~km~s$^{-1}$, between the carbon and hydrogen RRLs indicates that these lines arise in different media or volumes. This velocity offset was also observed by R97.

The CRRL optical depths observed here show a behavior that is consistent with cold, diffuse gas and are similar to those observed for the cool clouds along the line of sight to Cassiopeia~A \citep[e.g.][]{Oo17,Sa17,Sa17b,Ka98a,Pa89}. This is particularly clear from the emission to absorption transition being between 100 and 200~MHz \citep{Oo17} and also directly shows that the CRRL emission must be stimulated \citep[e.g.][R97]{Sa17a,Pa89}. \citet{Sa17b} has previously attempted to model the GC region using optical depth and linewidth ratios of C$\alpha$ and C$\beta$ lines. They find that models with \textit{T}$_{\rm{e}}\sim$20--60~K and \textit{n}$_{\rm{e}}\sim$0.04--0.1~cm$^{-3}$ are consistent with the observations. Our data are in agreement with this.

Having briefly discussed the CRRL measurements we will for the remainder of this study focus on the HRRLs and in particular on their line widths.

\subsection{Hydrogen RRLs}\label{s_discuss_hrrl}
Our HRRL measurements overlap with previous work around 145~MHz by APB90, 241 and 328~MHz by P78 and 328~MHz by R97. Comparing with these studies, our measurements are largely consistent in terms of \textit{v}$_{\rm{LSR}}$ velocity and FWHM line width. In terms of optical depth our detections are typically lower, except near 145~MHz where we agree with APB90. The APB90 measurements have a similar beam size as the EDA and hence the dominant part of the variations in calculated optical depths by different measurements can therefore be understood in terms of their beam sizes. Our observations with the EDA typically have a larger beam size and measure lower optical depths. This indicates that on degree scales the HRRL emitting gas suffers from beam dilution.

Above 80~MHz, our HRRL detections are consistent with a single velocity component at \textit{v}$_{\rm{LSR}}\sim$10~km~s$^{-1}$ with a FWHM$\sim$25~km~s$^{-1}$. This component has previously been referred to as the 0~km~s$^{-1}$ component by P78, but later studies find it to be closer to 10~km~s$^{-1}$ (e.g. R97). Between 105 and 325~MHz our optical depths are nearly constant. This supports previous conclusions that this low-frequency component of the HRRL emission is dominated by stimulated emission (e.g. P78; R97). Our H$\beta$ detection at 183~MHz furthermore indicates that the low-frequency HRRL emitting gas is not in LTE (see Sect.~\ref{s_ab_ratio}).

\subsection{HRRL line width}\label{s_width_hii}
Our detections with the EDA are the lowest frequency HRRL detections to date. Previously these were the 141.5 (\textit{n}=359) and 148.9~MHz (\textit{n}=353) detections by APB90. Below 80~MHz we find, for the first time, that the HRRL line width decreases from a FWHM of 25~km~s$^{-1}$ to only 12~km~s$^{-1}$. Fig~\ref{f_hrrl_mplt} shows that this change happens gradually between 190 and 80~MHz. Below 80~MHz the line is only marginally resolved by the EDA and as such we will treat this width as an upper limit. The decrease in line width could be explained by beam dilution if the 25~km~s$^{-1}$ line width component is less extended than the 12~km~s$^{-1}$ component. This would then be similar to the HRRL line profile at 1.6~GHz that shows a decrease in line width with increasing beam size (P78). 

At 73 and 63~MHz, where we observe the narrow 12~km~s$^{-1}$ line, our FWHM beam size is 6.7 and 7.8 deg. This is much larger than previous measurements and hence it is possible that the 25~km~s$^{-1}$ line width component is diluted to such an extent that only an even more extended component with a line width of 12~km~s$^{-1}$ component remains. Alternatively, it is also possible that the 25~km~s$^{-1}$ line width component consists of several, different velocity components of which only a component with a width of 12~km~s$^{-1}$ remains observable below 80~MHz. Higher spatial and spectral resolution observations below 100~MHz are necessary to test this.

Evidence for the existence of a narrow HRRL component towards the GC has previously been presented at 328~MHz by R97. In their average spectra, they observe a line profile that is consistent with a blend of a narrow 8~km~s$^{-1}$ and a broad 38~km~s$^{-1}$ component. They also show that this line profile is extended over at least 1~degree in longitude. Due to difficulties with RFI mitigation, R97 could not conclusively verify this narrow component at 328~MHz. Near 310~MHz, our line profiles are best described by a single component with a width of about 20~km~s$^{-1}$, but our spectra have lower signal to noise than R97.

For our narrow 12~km~s$^{-1}$ HRRL line width detections at 63 and 73~MHz we can rule out RFI, as we (i) detect the line in each of the two stacks independently and (ii) observe a gradual transition from a 25~km~s$^{-1}$ line width at 190 MHz to 12~km~s$^{-1}$ below 80~MHz.

\subsection{Cold H$^{+}$ at 63~MHz}\label{s_cold_hii}
In previous work, P78 and APB90 favor an interpretation where the origin of the low-frequency 25~km~s$^{-1}$ HRRL line width component is warm (\textit{T}$_{\rm{e}}\sim$7000~K), low density (\textit{n}$_{\rm{e}}\sim$5--10~cm$^{-3}$) gas from the outer envelopes of HII regions. This interpretation is based on the assumption that the measured line width arises due to a single, homogeneous velocity component. However, in the case that it consists of several, narrower components then the warm models no longer apply. For example, R97 showed that the maximally allowed temperature, considering only Doppler broadening, for their narrow line feature is 1380~K. Similarly narrow (3--10~km~s$^{-1}$) HRRL components have been found in detailed studies of cool, diffuse clouds at 226--408~MHz \citep*{Oo17,So10} and in HII regions near 1.4~GHz \citep*[e.g.][and references therein]{Pa77,Ro92}. In the latter case these narrow features are typically blended with much broader ($\sim$60~km~s$^{-1}$) line emission that dominates the total HRRL signal at GHz frequencies.   

Our measurements show that a narrow HRRL component exists along the line of sight to the GC and dominates the HRRL signal below 80~MHz. Considering the very large beam sizes involved, this emission must be extended over degree scales to remain detectable. It therefore seems likely that this gas is located relatively close to us. Considering only Doppler broadening we can place an upper limit of 3200~K for the temperature of the HRRL emitting gas. However, contrary to earlier studies our detections are at sufficiently low frequencies that pressure and radiation broadening also need to be considered. This enables us to not only constrain the temperature, but also the maximally allowed product of electron density and temperature for the emitting material. At the same time, the absence of significant line broadening observed here between 74 and 63~MHz implies that radiation and pressure broadening cannot dominate the overall line width at these frequencies. The strongest constraint on \textit{n}$_{\rm{e}}$ and \textit{T}$_{\rm{e}}$ is then obtained at the lowest frequency, i.e. at 63~MHz (\textit{n}=470). 

The long line of sight towards the GC has an integrated surface brightness temperature \textit{T}$_{\rm{R,100}}\sim$10000~K at 100~MHz \citep[e.g.][]{Do17}. Given the likely proximity of the gas we will here assume that it sees only a fraction of this radiation. We will therefore conservatively use a power law with $\lambda^{2.6}$ normalized at 100~MHz through \textit{T}$_{\rm{R,100}}$=2000~K to describe the radiation field experienced by the HRRL emitting gas. At 63~MHz this will produce only 1.8~km~s$^{-1}$ of broadening and this amount scales linearly with the value of \textit{T}$_{\rm{R,100}}$ \citep[e.g.][]{Sa17b,Sh75}. Higher \textit{T}$_{\rm{R,100}}$ values will lead to even more stringent upper constraints on \textit{T}$_{\rm{e}}$ and \textit{n}$_{\rm{e}}$, whereas lower \textit{T}$_{\rm{R,100}}$ values imply that radiation broadening will have a neglible effect on the line width.

\begin{figure*}
\mbox{
    \includegraphics[width=1.0\textwidth, angle=0]{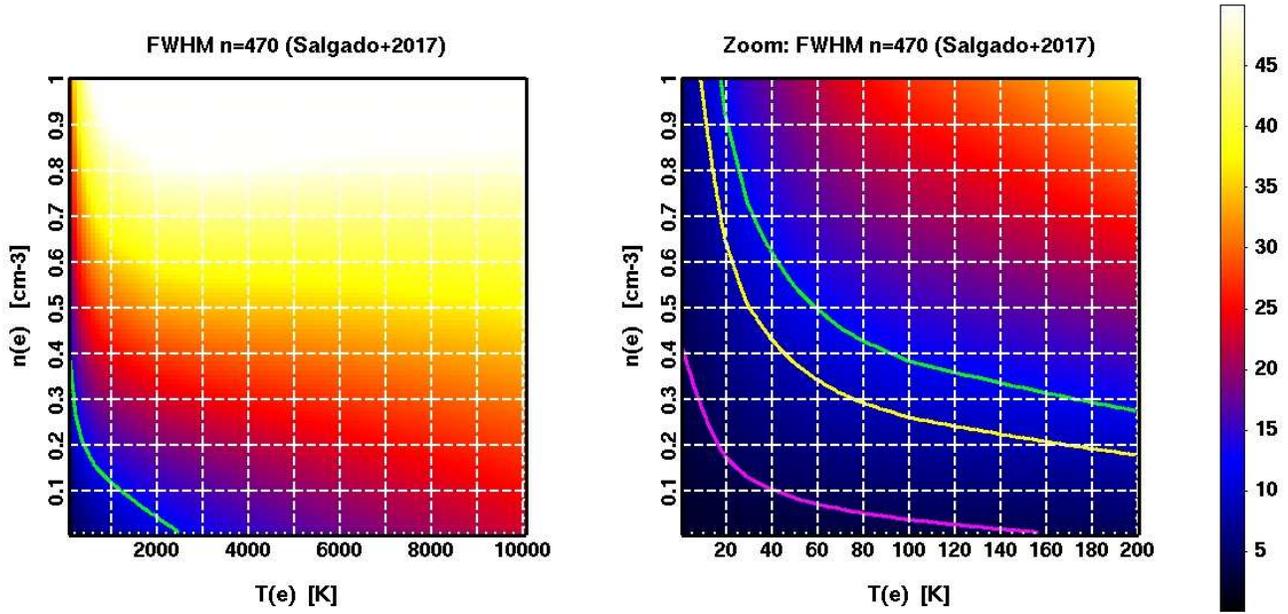}
}
  \vspace{0.5cm}
  \caption{Line broadening (color scale) in units of km~s$^{-1}$ as a function of gas temperature \textit{T}$_{\rm{e}}$ and density \textit{n}$_{\rm{e}}$ for a fixed Galactic, power-law ($\beta$=2.6) radiation field normalized at 100~MHz by \textit{T}$_{\rm{R,100}}$=2000~K. We consider Doppler, pressure and radiation broadening here following \citet{Sa17b}. (Left) Showing the line width for the full grid in \textit{T}$_{\rm{e}}$ and \textit{n}$_{\rm{e}}$. Here the solid green line shows the maximally allowed combination of pressure and temperature for the HRRL line width of 12~km~s$^{-1}$ at 63~MHz (\textit{n}=470). (Right) A zoom-in on the low temperature (0--200~K) part of the grid. The green (top) solid line is the same as in the left panel. The yellow (middle) and magenta (bottom) lines show the corresponding \textit{T}$_{\rm{e}}$, \textit{n}$_{\rm{e}}$ limits for line widths of 9 and 4~km~s$^{-1}$ respectively.}\label{f_fwhm_tn}
\end{figure*}

Combining Doppler and radiation broadening with pressure broadening (S17b) we constrain the maximally allowed product of \textit{T}$_{\rm{e}}$ and \textit{n}$_{\rm{e}}$ at 63~MHz (\textit{n}=470). This is shown as the green line in Fig.~\ref{f_fwhm_tn}. For \textit{n}$_{\rm{e}}\leq$0.2~cm$^{-3}$ and \textit{T}$_{\rm{e}}\gtrsim$1000~K Doppler broadening exceeds pressure broadening and drives the allowed densities down with increasing temperature. The green line is a true upper limit as the HRRL line width at 63 MHz is unresolved with the EDA. 

The recent RRL optical depth models of \citet{Sa17a} can be used to estimate the H$^{+}$ column density associated with our HRRL detection at 63~MHz. For \textit{T}$_{\rm{e}}$=10--150~K and \textit{n}$_{\rm{e}}$=0.01--0.11~cm$^{-3}$ a detailed grid was computed by \citet{Oo17}. For this range we require path lengths of L$_{\rm{HII}}$=0.5--50~pc and column densities N$_{\rm{HII}}$=2$\times$10$^{16}$--2$\times$10$^{18}$~cm$^{-2}$ to explain the observed optical depth. The largest path lengths are obtained for the lowest densities and highest temperatures. This also implies that temperatures significantly higher than about 1000~K are not valid solutions, as these would have unphysical path lengths. For example, the maximally allowed temperature for \textit{n}$_{\rm{e}}$=0.01~cm$^{-3}$ is \textit{T}$_{\rm{e}}$=2500~K and requires L$_{\rm{HII}}$=4~kpc and N$_{\rm{HII}}$=1$\times$10$^{20}$~cm$^{-2}$ to explain the optical depth at 63~MHz.

Recently there has been discussion in the literature about the different collision rates used for \textit{l}-changing transitions \citep*{Vr12,Gu16,Gu17,Vr17}. Here \textit{l} is angular momentum. In particular it has been shown that the semiclassical derived rates in \citet[][hereafter VOS12]{Vr12} underpredict the corresponding quantum mechanical rates for some l-changing collisions. The S17a models that compute the departure coefficients necessary for calculating the HRRL optical depth use the VOS12 semiclassical rates. We find that the differences in the HRRL departure coefficients computed using the VOS12 rates as compared to the coefficients computed using either the \citet*{Pe64} or the new \citet{Vr19} analytical expressions for the rates are not significant for n$\gtrsim$300 when considering the physical conditions of interest here. The \citet{Pe64} and the \citet{Vr19} rates both approximate the quantum mechanical rates well and an indepth comparison is provided in \citet{Vr19}.

As an example, we find for a homogeneous gas slab with \textit{T}$_{\rm{e}}$=40~K and \textit{n}$_{\rm{e}}$=0.1~cm$^{-3}$ (see Sect.~\ref{s_origin}) that the product of the departure coefficients (\textit{b}$_{n}\times\beta_{n}\propto\int\tau d\nu$) for n$\gtrsim$300 calculated with the three different rates agree with each other to within 1 percent. This can be understood as the \textit{l}-changing collision rates are known to primarily affect low to intermediate n values \citep[e.g.][]{Sa17a,Hu87}. We are therefore confident that the differences in the \textit{l}-changing collision rates discussed in the recent literature do not affect the results presented here. Furthermore, these rates do not affect the HRRL line width analysis presented in Sect.~\ref{s_width_hii}.

\subsection[]{Origin of the cold H$^{+}$ towards the GC}\label{s_origin}
In the previous section we showed that the narrow line width of the HRRL at 63~MHz can only be explained by a cold (partially) ionized gas. The thermal pressure of such a gas, for \textit{T}$\sim$10$^{2}$~K with a maximally allowed density \textit{n}$_{\rm{e}}$=0.4~cm$^{-3}$, is \textit{p}$\sim$\textit{n}$_{\rm{H}}\times$\textit{T}=10$^{2}\times$0.4/[\textit{n}$_{\rm{H^{+}}}$/\textit{n}$_{\rm{H}}$] in units K$\times$cm$^{-3}$. Here [\textit{n}$_{\rm{H^{+}}}$/\textit{n}$_{\rm{H}}$] is the hydrogen ionization fraction and we assume that \textit{n}$_{\rm{e}}$=\textit{n}$_{\rm{H^{+}}}$. Since the product of \textit{T}$_{\rm{e}}$ and \textit{n}$_{\rm{e}}$ is an upper limit, the ionization fraction is also an upper limit. For the above temperature and density we require that [\textit{n}$_{\rm{H^{+}}}$/\textit{n}$_{\rm{H}}$]$\leq$(2--10)$\times$10$^{-3}$ to remain within the range of typical pressures (3$\times$10$^{3}$ -- 2$\times$10$^{4}$ K$\times$cm$^{-3}$) measured in the Galactic ISM \citep*[e.g.][]{Oo17,Je11,Wo03}. 

With these conditions the HRRL emitting gas at 63~MHz would originate in the cold neutral medium (CNM), likely arising in the cold partially ionized skins of diffuse clouds in the Galactic disc \citep[e.g.][]{Oo17}. Requiring pressure balance with the ISM also rules out an origin for the HRRL emitting gas in either the warm ionized medium (WIM) or the warm neutral medium (WNM), as the associated thermal pressures would be too low.

A prominent, nearby cool cloud along the line of sight to the GC is the Riegel-Crutcher (RC) cloud \citep*{Ri72}. This cloud is located at at a distance of about 125~pc and spans at least 40 degrees in Galactic longitude (\textit{l}=+25 to \textit{l}=-15 deg) and 10 degrees in Galactic latitude (\textit{b}=-3 to \textit{b}=+7 deg). It has been observed in HI absorption \citep*[e.g.][]{Mc06} and various molecular transitions such as CO and OH \citep[e.g.][]{Ok98,Cr73}. The cold, atomic gas in this cloud is estimated to have a temperature of about 40~K, a column density of N(HI)=(1--4)$\times$10$^{20}$ and a total line of sight thickness between 1 and 5~pc. The HI observations show that the RC cloud consists of many thin, $\sim$0.1~pc, strands \citep{Mc06}. This type of sub-structure may be responsible for the velocity offset measured here between HRRL and CRRL. 

At \textit{T}$_{\rm{e}}$=40~K the HRRL line width constrains \textit{n}$_{\rm{e}}\leq$0.6cm$^{-3}$. However, the models by \citet{Sa17a} require that \textit{n}$_{\rm{e}}\leq$0.1~cm$^{-3}$ for HRRL to be observed in emission. For this combination we find L$_{HII}$=1~pc, N$_{HII}$=3$\times$10$^{17}$cm$^{-2}$, [n$_{H^{+}}$/n$_{H}$]$\sim$1$\times$10$^{-3}$ and p$_{th}\sim$4000~K~cm$^{-3}$. This path length and pressure are consistent with the HI measurements for the RC cloud. The low-frequency CRRL emission near the GC has previously been attributed to the RC cloud and found to be consistent with the physical conditions estimated for the HI absorbing gas \citep{Ro11,Sa17b}.

The temperature and density limits derived here for the HRRL emission at 63~MHz therefore imply that both the HRRL and CRRL emission could be associated with the RC cloud. In particular they could arise in different layers of the partially ionized envelope of the cloud. This would then be similar to the low-frequency RRL emission observed from the cool clouds along the line of sight to Cassiopeia~A \citep[e.g.][]{Oo17,Sa18}.

A cold, ionized plasma will also contribute to low-frequency continuum absorption via free-free processes \citep{Do17,Ka95,Is90,Is92}. Considering the physical conditions for the RC cloud, the HRRL emission at 63~MHz does not cause significant absorption. However, if the broader, FWHM$\sim$25~km~s$^{-1}$ HRRL emission is a blend of multiple cold, diffuse (to avoid beam dilution) gas components, then these must have a higher density to avoid detection below 80~MHz and could then account for the observed continuum absorption. Additional detailed broad-band, high spatial and spectral resolution observations with e.g. LOFAR, MWA and the future SKA1-Low, are necessary to further investigate this.

\subsection[]{Future SKA1-Low observations}\label{s_ska_future}
The scientific importance of sensitive, low-frequency RRL surveys to probe the the diffuse CNM, with the future SKA1-Low, was outlined in \citet{Oo15}. There we focused on how CRRL, in combination with HRRL and HI 21~cm measurements, allow us to derive important physical properties of the diffuse CNM that are not easily obtained via other methods. In particular we showed how the thermal pressure, carbon abundance, warm to cold fraction of the HI~21~cm signal and the hydrogen ionization rate can be obtained, and specific examples were given for the GC region and the Magellanic Clouds.

In \citet{Oo17} and \citet{Sa18} we showed that low-frequency CRRLs trace cool, diffuse clouds in the CNM and we found that the physical conditions of CRRL emitting gas are consistent with CO-dark gas. This would, potentially, make the CRRLs an important probe of CO-dark gas. Low-frequency HRRLs appear to be related to the corresponding CRRLs and their characteristics show that they must also arise in cool, partially ionized gas (this work). However, we find that the HRRLs do not match the CRRLs exactly, in either the derived physical conditions \citep{Oo17} or in velocity space (this work). To make progress in understanding the low-frequency HRRL emitting gas we require the following: (i) better spatial resolution to map its morphology and relate it to other known gas phases, and (ii) more sensitive, high spectral resolution, wide-band observations to derive its physical conditions.

The future SKA1-Low can meet these requirements and will enable us to determine the thermal balance, chemical enrichment, and ionization rate of the cold, diffuse medium from degree scales down to scales corresponding to individual clouds and filaments in our Galaxy. Table~1 in \citet{Oo15} provides 5$\sigma$ peak optical depth limits for observing extended RRL emission with SKA1-Low when using the core array (2~km effective baseline) with 1~kHz channels and stacking the lines in groups of 9.\footnote{Please note that the numbers in \citet{Oo15} refer to an early version of the SKA1-Low design. The design for SKA1-Low has not been finalized and as such the detailed sensitivity across the frequency band is not yet known.}
The corresponding column density limits for observing low-frequency HRRL emitting gas with SKA1-Low can then be calculated for a given set of physical conditions. As an example we will here use a homogenous gas slab, observed along the line of sight to the GC, with physical conditions similar to those discussed for the RC cloud in Sect.~\ref{s_origin}, i.e. $T_{\rm{e}}$=40~K, $n_{\rm{e}}$=0.1~cm$^{-3}$, $T_{\rm{R,100}}$=2000~K and FWHM=12~km~s$^{-1}$. Invoking the S17a models and Table~1 in \citet{Oo15}, SKA1-Low will be able to detect H$^{+}$ column densities in the range (4--20)$\times$10$^{15}$~cm$^{-2}$ at the 5$\sigma$ line peak level and with an angular resolution between 1.5 and 8.6~arcmin. At the distance of the RC cloud this angular resolution corresponds to sub-pc scale spatial resolution. Considering hydrogen ionization fractions of 10$^{-4}$ to 10$^{-3}$ this corresponds to a neutral gas column density of about 10$^{19}$ to 10$^{20}$~cm$^{-2}$.

The sensitivity of SKA1-Low for RRLs decreases at the high-frequency end of the band and below 110~MHz. The latter is due to a steep drop-off in the bandpass below 110~MHz. Towards the higher frequencies it is a combination of a slow bandpass drop-off and a decreasing Galactic background radiation field. Broadband sensitivity is very important to capture the distinctive features of RRL models, e.g. the emission to absorption transition and maximum emission turnover \citep{Oo17}. However, currently the biggest issue for observing low-frequency RRLs from cool, diffuse gas with SKA1-Low is the recently planned decrease in the numbers of channels from 262144 to 65536. As shown here, and in \citet{Oo17,Oo15}, these low-frequency RRL observations require a channel width of at most 1~kHz to obtain the minimally necessary velocity resolution, i.e. 2 to 12~km$^{-1}$ across the 50--350~MHz range. The reduction in the number of channels for SKA1-Low therefore has important consequences for RRL measurements in that the available bandwidth is effectively reduced from 300~MHz to 4 zoom windows covering 16~MHz each. This setup would then allow for only 21~\% of the available RRLs to be observed instantaneously and thus a factor 5 loss in observing time. Furthermore, to clearly resolve the HRRL lines, as observed here, even higher spectral resolution would be necessary at the lowest frequencies.

\section[]{Conclusions}\label{s_conclus}
We have presented 0--328~MHz spectroscopic observations with the EDA towards the GC region. Useful data were obtained in 30--325~MHz range. We find that about 10~percent of our data is affected by RFI. This percentage can be lowered by observing with higher spectral resolution, as shown by LOFAR \citep[e.g.][]{Oo17,Sa17}. The antenna response, as reflected in the bandpass for the EDA, shows the expected decrease in sensitivity below 60~MHz and above 180~MHz.

With the EDA we have detected CRRLs between 39 and 325~MHz and HRRLs between 59 and 325~MHz respectively towards the GC region. For the HRRLs our observations are the first detection of this line below 145~MHz and reveal a previously unknown cold, diffuse, ionized gas component. We summarize our main conclusions below.

\begin{itemize}
\item The peak HRRL emission is located at LSR velocities +10 to +15~km~s$^{-1}$, and is located at a velocity 3-4~km~s$^{-1}$ higher than CRRL over the entire frequency range. Both HRRL and CRRL shift by about 5~km~s$^{-1}$, towards higher LSR velocities, with decreasing frequency and increasing beam size.
\\
\item The slowly varying optical depth of HRRL and the emission to absorption transition of CRRL show that the RRL emitting gas towards the GC is dominated by stimulated emission.
\\
\item The $\alpha$ to $\beta$ line ratios for both HRRL and CRRL imply that the low-frequency RRL emitting gas is not in LTE.
\\
\item The EDA CRRL measurements are consistent with previous work by E95, R97, and S17b showing that the CRRL emission most likely originates in cool, diffuse cloud emission that is part of the CNM. 
\\
\item Below 80~MHz we find a decrease in HRRL line width from a FWHM of 25~km~s$^{-1}$ to $\leq$12~km~s$^{-1}$. This decrease, in combination with the large beam size, reveals that a cold, diffuse H$^{+}$ gas exists on degree scales towards the GC region and dominates at the lowest frequencies.
\\
\item The optical depth of the cold, narrow (FWHM=12~km~s$^{-1}$) H$^{+}$ emission at 63~MHz does not cause significant free-free absorption. The broader 25~km~s$^{-1}$ component, if cold, would be able to explain the observed continuum absorption at low frequencies.
\end{itemize}

The narrow, weak nature of the HRRL emission observed below 80~MHz explains why it has been difficult to measure these lines in the past. These HRRLs are likely part of the same CNM component as the CRRL, albeit that they do not arise in exactly the same volume. The presence of low-frequency HRRL and CRRL associated with cold, diffuse gas is consistent with recent models \citep[e.g.][]{Sa17a}. A possible origin for the RRL emission observed here is the nearby RC cloud, as its promixity, size and the background radiation field create a particularly favorable situation for detection with the EDA.

Future observations with SKA1-Low will enable us to determine the morphology and physical conditions of the low-frequency HRRL emitting gas discovered here. However, improved broadband sensitivity with (sub-)kHz spectral resolution is desirable in order for SKA1-Low to efficiently observe these low-frequency RRLs over large scales in the Galaxy. If achieved, this would enable detailed RRL modeling constraining the physical conditions of the media in which they arise on arcsec to arcmin scales \citep[e.g.][]{Oo17,Oo15, Sa18,Sa17,Sa17a}.

\section*{Acknowledgments}
J.B.R.O. acknowledges financial support from NWO Top LOFAR-CRRL project, project No. 614.001.351. E.L.A. would like to thank the ASTRON/JIVE summer student programme for a fellowship in 2017. This scientific work makes use of the Murchison Radioastronomy Observatory, operated by CSIRO. We acknowledge the Wajarri Yamatji people as the traditional owners of the Observatory site.

%



\clearpage

\newpage

\begin{table*}
 \centering
  \begin{tabular}{|l|r|} \hline
  Parameter                & EDA  	 \\ \hline
  Field center RA (J2000)  & 17h45m40s   \\
  Field center DEC (J2000) & -29d00m28s  \\
  Oberving date            & 13-07-2017  \\
  Oberving start (UT)      & 14:54:53   \\
  Oberving end   (UT)      & 18:15:10   \\
  Total on-source time     & 03h20min    \\
  Frequency range          & 0--327.68 MHz  \\
  Number of channels       & 262144      \\
  Channel width            & 1.25 kHz    \\ \hline
  \end{tabular}
 \caption[]{Details of the EDA observation. For the frequency range please note that only the range 30--325~MHz provides useful data. Over the 30--325~MHz range the spatial resolution changes from 1.5 to 16.4~deg and the spectral resolution changes from 2.3 to 25~km~s$^{-1}$}.\label{t_obs}
\end{table*}

\begin{table*}
 \centering
  \begin{tabular}{|l|l|l|r|r|r|} \hline
  C$\alpha$ transition    	& Frequency	& FWHM$_{beam}$ & $\int\tau$d$\nu$		& \textit{v}$_{\rm{LSR}}$	& FWHM$_{line}$ \\ \hline  
  \textit{n}-range (\#lines)	& [MHz]		& [deg]		& [Hz]				& [km~s$^{-1}$]		& [km~s$^{-1}$] \\ \hline  
  272--281 (8)			& 310.1		& 1.6		& -2.82~$\pm$~0.78		& 5.4~$\pm$~2.4		& 17.4~$\pm$~5.6 \\ 
  288--289 (2)			& 272.6		& 1.8		& $\leq$2.24			& ---			& [20] \\ 
  302--305 (4)			& 234.2		& 2.1		& $\leq$1.11 			& 5.7~$\pm$~1.3		& [20] \\ 
  306--318 (13)			& 216.1		& 2.3		& -2.61~$\pm$~0.53		& 6.2~$\pm$~1.7		& 17.2~$\pm$~4.0 \\ 
  319--333 (15)			& 189.6		& 2.6		& $\leq$0.47			& ---			& [20] \\ 
  334--341 (8)			& 170.5		& 2.9		& $\leq$0.32			& ---			& [20] \\ 
  343--360 (18)			& 151.4		& 3.2		& $\leq$0.21			& ---			& [30] \\ 
  378--397 (20)			& 113.0		& 4.3		& 5.27~$\pm$~0.40		& 7.3~$\pm$~1.4		& 38.3~$\pm$~3.4 \\ 
  411--434 (24)			&  87.3		& 5.6		& 5.96~$\pm$~0.33		& 10.5~$\pm$~0.8	& 30.5~$\pm$~2.0 \\ 
  435--460 (26)			&  73.5		& 6.7		& 7.36~$\pm$~0.25		& 9.6~$\pm$~0.5		& 32.3~$\pm$~1.2 \\ 
  461--480 (20)			&  63.1		& 7.8		& 6.89~$\pm$~0.24		& 10.7~$\pm$~0.5	& 30.9~$\pm$~1.2 \\ 
  481--500 (20)			&  55.7		& 8.8		& 6.54~$\pm$~0.31		& 11.7~$\pm$~0.7	& 29.1~$\pm$~1.6 \\ 
  501--520 (20)			&  49.4		& 9.9		& 7.93~$\pm$~0.36		& 10.9~$\pm$~1.0	& 44.7~$\pm$~2.4 \\ 
  521--540 (20)			&  44.0		& 11.2		& 5.51~$\pm$~0.29		& 7.8~$\pm$~1.1		& 42.4~$\pm$~2.6 \\ 
  541--550 (10)			&  40.4		& 12.2		& 6.14~$\pm$~0.52		& 5.4~$\pm$~3.0		& 56.5~$\pm$~??? \\ \hline 
  \end{tabular}
 \caption[]{Cn$\alpha$ stacked RRL measurements. The integrated line optical depth $\int\tau$d$\nu$, local standard of rest velocity \textit{v}$_{\rm{LSR}}$ and full width at half maximum line width FWHM$_{line}$ are obtained from single Gaussian fits to the observed line profiles; see Fig.~\ref{f_crrl_mplt}. The line optical depth $\tau$ is defined as -$T_{\rm{line}}/T_{\rm{cont}}$, with $T_{\rm{line}}$ the line temperature and $T_{\rm{cont}}$ the continuum temperature. The first column shows the quantum range considered and the values in the parentheses provide the number of lines stacked. The upper limits for non-detections are given in units of optical depth at the 3$\sigma$ level and are integrated over the expected line width that is given in square brackets in the FWHM column.}\label{t_cna_rrl_stacked}
\end{table*}

\begin{table*}
 \centering
  \begin{tabular}{|l|l|l|r|r|r|} \hline
  C$\beta$ transition  		& Frequency	& FWHM$_{beam}$ & $\int\tau$d$\nu$		& \textit{v}$_{\rm{LSR}}$	& FWHM$_{line}$ \\ \hline  
  \textit{n}-range (\#lines)	& [MHz]		& [deg]		& [Hz]				& [km~s$^{-1}$]		& [km~s$^{-1}$] \\ \hline  
  342--354 (13)			& 310.1		& 1.6		& $\leq$0.48			& ---			& --- \\ 
  387--453 (67)			& 183.0		& 2.7		& 2.64~$\pm$~0.32		& 9.8~$\pm$~1.8		& 30.8~$\pm$~4.3 \\ 
  476--500 (25)			& 112.9		& 4.3		& 2.83~$\pm$~0.31		& 9.7~$\pm$~1.6		& 28.7~$\pm$~3.7 \\ 
  520--605 (86)			& 76.1		& 6.5		& 3.32~$\pm$~0.13		& 10.6~$\pm$~0.6	& 33.7~$\pm$~1.5 \\ 
  606--696 (91)			& 48.8		& 10.1		& 4.19~$\pm$~0.25		& 9.1~$\pm$~2.5		& 85.6~$\pm$~5.9 \\ \hline 
  \end{tabular}
 \caption[]{Cn$\beta$ stacked RRL measurements. The $\int\tau$d$\nu$, \textit{v}$_{\rm{LSR}}$ and FWHM$_{line}$ are obtained from single Gaussian fits to the observed line profiles. Dee Fig.~\ref{f_chbeta} and Table~\ref{t_cna_rrl_stacked} for more information.}\label{t_cnb_rrl_stacked}
\end{table*}

\begin{table*}
 \centering
  \begin{tabular}{|l|l|l|r|r|r|} \hline
  H$\alpha$ transition (bin)   	& Frequency	& FWHM$_{beam}$ & $\int\tau$d$\nu$		& \textit{v}$_{\rm{LSR}}$	& FWHM$_{line}$ \\ \hline  
  \textit{n}-range (\#lines)	& [MHz]		& [deg]		& [Hz]				& [km~s$^{-1}$]		& [km~s$^{-1}$] \\ \hline  
  272--281 (8)			& 309.9		& 1.6		& -5.39~$\pm$~0.84		&  9.6~$\pm$~0.7	& 19.4~$\pm$~3.5 \\ 
  288--289 (2)			& 272.4		& 1.8		& -8.24~$\pm$~1.71		& 15.8~$\pm$~2.1	& 20.4~$\pm$~4.9 \\ 
  302--305 (4)			& 234.1		& 2.1		& -8.02~$\pm$~1.32		&  6.7~$\pm$~1.6	& 19.5~$\pm$~3.7 \\ 
  306--318 (13)			& 216.0		& 2.3		& -8.75~$\pm$~0.54		& 10.2~$\pm$~0.7	& 23.8~$\pm$~1.7 \\ 
  319--333 (15)			& 189.5		& 2.6		& -7.81~$\pm$~0.59		&  9.4~$\pm$~1.0	& 26.8~$\pm$~2.3 \\ 
  334--341 (8)			& 170.4		& 2.9		& -6.61~$\pm$~0.47		& 11.6~$\pm$~0.8	& 24.2~$\pm$~2.0 \\ 
  343--360 (18)			& 151.3		& 3.2		& -5.68~$\pm$~0.40		&  8.1~$\pm$~0.9	& 24.8~$\pm$~2.0 \\ 
  378--397 (20)			& 113.0		& 4.3		& -3.67~$\pm$~0.40		&  9.2~$\pm$~1.5	& 29.3~$\pm$~3.7 \\ 
  411--434 (24)			&  87.3		& 5.6		& -2.61~$\pm$~0.33		& 12.2~$\pm$~2.6	& 41.6~$\pm$~6.1 \\ 
  435--460 (26)			&  73.5		& 6.7		& -0.98~$\pm$~0.16		& 12.9~$\pm$~0.9	& 11.5~$\pm$~2.2 \\ 
  461--480 (20)			&  63.1		& 7.8		& -0.56~$\pm$~0.15		& 15.8~$\pm$~1.6	& 11.9~$\pm$~3.7 \\ 
  481--500 (20)			&  55.7		& 8.8		& $\leq$0.056			& ---			& [12.0] 	\\ \hline 
  \end{tabular}
 \caption[]{Hn$\alpha$ stacked RRL measurements. The $\int\tau$d$\nu$, \textit{v}$_{\rm{LSR}}$ and FWHM$_{line}$ are obtained from single Gaussian fits to the observed line profiles. See Fig.~\ref{f_hrrl_mplt} and Table~\ref{t_cna_rrl_stacked} for more information.}\label{t_hna_rrl_stacked}
\end{table*}

\begin{table*}
 \centering
  \begin{tabular}{|l|l|l|r|r|r|} \hline
  H$\beta$ transition  		& Frequency	& FWHM$_{beam}$ & $\int\tau$d$\nu$		& \textit{v}$_{\rm{LSR}}$	& FWHM$_{line}$ \\ \hline  
  \textit{n}-range (\#lines)	& [MHz]		& [deg]		& [Hz]				& [km~s$^{-1}$]		& [km~s$^{-1}$] \\ \hline  
  342--354 (13)			& 310.0		& 1.6		& $\leq$0.51			& ---			& [20] \\ 
  387--453 (67)			& 182.9		& 2.7		& 0.80~$\pm$~0.24		& 159.7~$\pm$3.2	& 22.4~$\pm$~7.6 \\ 
  476--500 (25)			& 112.9		& 4.3		& $\leq$0.14			& ---			& [20] \\ 
  520--605 (86)			& 76.1		& 6.5		& $\leq$0.056			& ---			& [20] \\ 
  606--696 (91)			& 48.8		& 10.1		& $\leq$0.046			& ---			& [12] \\ \hline 
  \end{tabular}
 \caption[]{Hn$\beta$ stacked RRL measurements. The $\int\tau$d$\nu$, \textit{v}$_{\rm{LSR}}$ and FWHM$_{line}$ are obtained from single Gaussian fits to the observed line profiles. See Fig.~\ref{f_chbeta} and Table~\ref{t_cna_rrl_stacked} for more information.}\label{t_hnb_rrl_stacked}
\end{table*}




\clearpage


\bsp

\label{lastpage}

\end{document}